\documentclass[conference,letterpaper]{IEEEtran}

\usepackage[utf8]{inputenc} 
\usepackage[T1]{fontenc}
\usepackage{url}
\usepackage{ifthen}
\usepackage{cite}
\usepackage[cmex10]{amsmath} 

\usepackage{graphicx}
\usepackage{overpic}
\usepackage{amsmath,amssymb,amsfonts}
\usepackage{textcomp}
\usepackage{array}
\usepackage{stfloats}
\usepackage{verbatim}
\usepackage{xcolor}
\usepackage{algorithmic}
\usepackage{setspace}
\usepackage{float}
\usepackage{caption}

\usepackage[caption=false,font=normalsize,labelfont=sf,textfont=sf]{subfig}
\begin{document}
  \newtheorem{definition}{\it Definition}
	\newtheorem{theorem}{\bf Theorem}
	\newtheorem{lemma}{\it Lemma}
	\newtheorem{corollary}{\it Corollary}
	\newtheorem{remark}{\it Remark}
	\newtheorem{example}{\it Example}
	\newtheorem{case}{\bf Case Study}
	\newtheorem{assumption}{\it Assumption}
	\newtheorem{property}{\it Property}
	\newtheorem{proposition}{\it Proposition}
	\newtheorem{observation}{\it Observation}

  \newcommand{\hP}[1]{{\boldsymbol h}_{{#1}{\bullet}}}
	\newcommand{\hS}[1]{{\boldsymbol h}_{{\bullet}{#1}}}
	
	\newcommand{\ba}{\boldsymbol{a}}
	\newcommand{\baq}{\overline{q}}
	\newcommand{\bA}{\boldsymbol{A}}
	\newcommand{\bb}{\boldsymbol{b}}
	\newcommand{\bB}{\boldsymbol{B}}
	\newcommand{\bc}{\boldsymbol{c}}
	\newcommand{\bcO}{\boldsymbol{\cal O}}
	\newcommand{\bh}{\boldsymbol{h}}
	\newcommand{\bH}{\boldsymbol{H}}
	\newcommand{\bl}{\boldsymbol{l}}
	\newcommand{\bm}{\boldsymbol{m}}
	\newcommand{\bn}{\boldsymbol{n}}
	\newcommand{\bo}{\boldsymbol{o}}
	\newcommand{\bO}{\boldsymbol{O}}
	\newcommand{\bp}{\boldsymbol{p}}
	\newcommand{\bq}{\boldsymbol{q}}
	\newcommand{\bR}{\boldsymbol{R}}
	\newcommand{\bs}{\boldsymbol{s}}
	\newcommand{\bS}{\boldsymbol{S}}
	\newcommand{\bT}{\boldsymbol{T}}
	\newcommand{\bw}{\boldsymbol{w}}
	
	\newcommand{\balpha}{\boldsymbol{\alpha}}
	\newcommand{\bbeta}{\boldsymbol{\beta}}
	\newcommand{\bOmega}{\boldsymbol{\Omega}}
	\newcommand{\bTheta}{\boldsymbol{\Theta}}
	\newcommand{\bphi}{\boldsymbol{\phi}}
	\newcommand{\btheta}{\boldsymbol{\theta}}
	\newcommand{\bvarpi}{\boldsymbol{\varpi}}
	\newcommand{\bpi}{\boldsymbol{\pi}}
	\newcommand{\bpsi}{\boldsymbol{\psi}}
	\newcommand{\bxi}{\boldsymbol{\xi}}
	\newcommand{\bx}{\boldsymbol{x}}
	\newcommand{\by}{\boldsymbol{y}}
	
	\newcommand{\cA}{{\cal A}}
	\newcommand{\bcA}{\boldsymbol{\cal A}}
	\newcommand{\cB}{{\cal B}}
	\newcommand{\cE}{{\cal E}}
	\newcommand{\cG}{{\cal G}}
	\newcommand{\cH}{{\cal H}}
	\newcommand{\bcH}{\boldsymbol {\cal H}}
	\newcommand{\cK}{{\cal K}}
	\newcommand{\cO}{{\cal O}}
	\newcommand{\cR}{{\cal R}}
	\newcommand{\cS}{{\cal S}}
	\newcommand{\dcS}{\ddot{{\cal S}}}
	\newcommand{\ds}{\ddot{{s}}}
	\newcommand{\cT}{{\cal T}}
	\newcommand{\cU}{{\cal U}}
	\newcommand{\wt}[1]{\widetilde{#1}}

	\newcommand{\mA}{\mathbb{A}}
	\newcommand{\mE}{\mathbb{E}}
	\newcommand{\mG}{\mathbb{G}}
	\newcommand{\mR}{\mathbb{R}}
	\newcommand{\mS}{\mathbb{S}}
	\newcommand{\mU}{\mathbb{U}}
	\newcommand{\mV}{\mathbb{V}}
	\newcommand{\mW}{\mathbb{W}}
	
	\newcommand{\uq}{\underline{q}}
	\newcommand{\ubq}{\underline{\boldsymbol q}}
	
	\newcommand{\red}[1]{\textcolor[rgb]{1,0,0}{#1}}
	\newcommand{\gre}[1]{\textcolor[rgb]{0,1,0}{#1}}
	\newcommand{\blu}[1]{\textcolor[rgb]{0,0,1}{#1}}

\title{Rateless Stochastic Coding for Delay-Constrained Semantic Communication} 


\author{%
  \IEEEauthorblockN{Cheng Peng}
  \IEEEauthorblockA{Huazhong University of Science and Technology\\
                    Wuhan, China}
  \IEEEauthorblockN{Yong Xiao}
  \IEEEauthorblockA{Huazhong University of Science and Technology\\ 
                    Wuhan, China}
  \and

  \IEEEauthorblockN{Rulong Wang}
  \IEEEauthorblockA{Huazhong University of Science and Technology\\ 
                   Wuhan, China}
  
}


\maketitle


\begin{abstract}
We consider the problem of joint source-channel coding for semantic communication from a rateless perspective, the purpose of which is to settle the balance between reliability (distortion/perception) and effectiveness (rate) of transmission over uncertain channels. 
In particular, we propose a more general communication objective that minimizes the perceptual distance by incorporating a semantic-level reconstruction objective in addition to the conventional pixel-level reconstruction objective.
Based on the proposed objective, we then propose a novel JSCC coding scheme called rateless stochastic coding (RSC) by introducing a generative decoder and dithered quantization.
The coding scheme enables reconstruction based on both distortion and perception metrics through rateless transmission.
Extensive experiments demonstrate that the proposed RSC can achieve variable rates of transmission maintaining an excellent trade-off between distortion and perception. 
\end{abstract}

\section{Introduction}

Amid the ongoing global revolution in the information industry, with a focus on 5G, 6G, and artificial intelligence technologies, numerous human-oriented intelligent services and applications such as virtual reality, augmented reality, tactile Internet, and autonomous driving are constantly being researched and deployed, highlighting the increasing importance of information semantics. This motivates a new communication paradigm, known as semantic communication, which concentrates on identifying, delivering, and utilizing critical meanings in messages during communication\cite{shi2021semantic}.

The concept of semantic communication was originally introduced through Weaver's proposition of a three-tiered communication problem that involves the "technical problem", the "semantic problem", and the "effectiveness problem"\cite{weaver1953recent}. According to Weaver, the semantic problem is mainly focused on the general interpretation of the received information. In other words, the key to addressing semantic issues is to extract the statistical semantic characteristics of the transmitted information. This also means that the typical symbol-level distortion metric employed in conventional communication is insufficient for evaluating the semantic distance. Recent research in generative modeling suggests that the divergence between data distributions can effectively represent the semantic perceptual quality. These studies offer a fresh perspective for assessing what Weaver refers to as statistical semantic characteristics.

Coding schemes in semantic communication face two challenges, the first is how to adaptively meet the individualized communication needs of different scenarios in terms of reliability and latency, and the second is the trade-off between protecting the semantic feature and symbolic accuracy of information during lossy encoding and the transmission of unknown channel. Coupled with feedback from the receiver, incremental redundancy hybrid automated repeat request (IR-HARQ) techniques achieve communication under various reliability constraints\cite{sesia2004incremental}. Rateless codes, which can be considered a form of continuous IR-HARQ\cite{castura2007rateless}, exhibit the capability to adapt to diverse channel conditions, achieving an effective balance between reliability and latency through dynamic coding rate adjustments. On the other hand, recent studies in neural compression, both simulated and theoretical, have illustrated a trade-off between divergence-based semantic perceptual quality and symbol-level distortion\cite{blau2019rethinking}\cite{zhang2021universal}. Additionally, for low transmission rates with imperfect channels, ensuring an exact level of signal distortion is highly challenging while there is potential for enhancing visual perceptual quality. For instance, the generation of high-visual-quality images by generative models can be viewed as a transmission process with a rate of zero. Therefore, incorporating the properties of rateless codes into the coding of semantic communication is a viable approach to achieving a flexible trade-off between distortion, perceptual quality, and latency via adaptive coding rate control. 

 

We summarize the key contributions of this paper as follows:
\begin{itemize}
    \item \emph{General optimizing objective of JSCC for semantic communication}: We propose a general optimizing objective of JSCC for semantic communication under delay constrains, in which not only minimum distortion is needed to achieve pixel-level reconstruction in the classical scheme, but also semantic-level reconstruction by minimizing perception.
    \item  \emph{Novel rateless stochastic coding scheme}: We propose RSC, a novel rateless stochastic coding scheme for minimizing the proposed objective, where MSE was used to measure distortion and Wasserstein distance to measure perceived distance, respectively.  In addition, we employ a tail drop operation on the encoded feature in order to achieve rateless coding.
    \item  \emph{Extensive experimental results}: We evaluate the performance of our proposed scheme by simulating the encoding and transmission over Gaussian channels on two standard image datasets. Compared to the traditional AE-based JSCC scheme, our coding scheme achieves better rate-distortion performance, especially at low coding rates. Additionally, our scheme achieves superior semantic visual perception which benefits the receiver in completing the artifact-free reconstruction of image data..
\end{itemize}

\section{Related work}

\subsection{Divergence-based Perception}
In recent years, numerous studies have explored lossy compression that incorporates visual perceptual constraints. According to Blau and Michaeli\cite{blau2018perception}, decreasing symbol-level distortion during data reconstruction does not always enhance visual perceptual quality. In contrast, compression algorithms that cater to visual perceptual constraints often perform inferior on distortion metrics, such as PSNR and SSIM. Subsequently, they proposed and proved a theory about the trade-off between distortion, perception, and rate under Bernoulli source\cite{blau2019rethinking}. Zhang et al. further demonstrated that the above properties also exist under Gaussian sources with specific distortion and perception metrics\cite{zhang2021universal}, Additionally, Wagner et al. developed a coding theory related to this trade-off and proved its reachability\cite{theis2021coding}. Theis et al. presented a coding example to demonstrate that by incorporating common randomness into the encoder and decoder, coding performance can be enhanced while accounting for perceptual quality\cite{theis2021advantages}. In other words, the stochastic encoders may outperform deterministic encoders when considering perceptual quality. Wagner et al. subsequently investigated the influence of varying levels of common randomness on the trade-off among rate, distortion, and perception\cite{wagner2022rate}. Inspired by these works, we introduce common randomness to construct a stochastic encoder/decoder in our coding scheme.

\subsection{Rateless code}
Rateless codes can adapt to various channel conditions by dynamically adjusting the coding rate. Therefore, they are commonly employed in erasure channels, time-varying channels, and broadcast channels. The rateless code is typically formulated as a streaming code with the capability to generate coded symbols indefinitely, ceasing transmission only upon the receiver's recovery of a single-bit ACK. As a class of classic rateless coding schemes, the fountain codes such as LT codes \cite{luby2002lt}and raptor codes\cite{shokrollahi2006raptor}, which were initially designed as a coding scheme to achieve the capacity of binary erasure channels (BEC), refined through various enhancements, also exhibit superior performance in Gaussian noise channels and wireless fading channels\cite{palanki2004rateless}\cite{castura2006rateless}. 
However, the rateless codes mentioned above use infinitely long codewords for bitstream transmission. Although they generally have a better expected effective length than fixed-rate codes, they lack theoretical reliability functions in scenarios with delay constraints. Draper et al. investigated the rateless code under arbitrary time-varying channels and provided an upper bound of error probability related to the decoding time. Blits derived the achievable rate of rateless codes through maximum posterior probability decoding under finite source message sets. It is evident from these works that in the finite coding regime, the error probability cannot converge to zero. This means that the channel does not guarantee reliable transmission, and therefore the separation theorem cannot be used to directly characterize rateless codes under distortion and perceptual constraints, Therefore, We model the problem as JSCC and present the achievable capacity region for rate-distortion-perception based on the dispersion proposed by Verdu. This result may provide theoretical support for Rateless JSCC under delay constraints.

\section{System Model and Problem Formulation}
\subsection{System Model}
\begin{figure}
		\centering
		\includegraphics[width=0.4\textwidth]{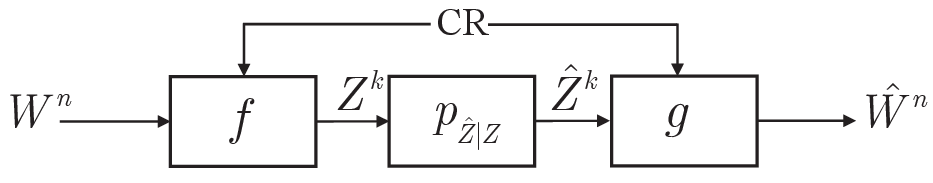}
		\caption{Illustration of a semantic communication system with CR over a noisy communication channel.}
		\label{f:JSCC_model1}
	\end{figure}

We consider the end-to-end semantic communication system introduced in \cite{chai2023rate} as shown in Fig. \ref{f:JSCC_model1}. 
        However, it can be noted that we take into account the limitation of delay in channel transmission and remove the side information setting.
	
In this system, the encoder can obtain a limited number $n$ of indirect observations of the semantic information source, denoted by $W^n$. The transmitter will then compress the indirect observations to $Z^k$ using a joint source-channel encoder $f$. Due to the limited channel capacity and maximum latency $K$, The signals sent to the channel by the encoder contain only $k\le K$ valid sequences, denoted as $Z^k$.  After transmission, the joint source-channel decoder $g$ decodes input $\hat{Z}^k$ to realize reconstruction $\hat{W}^n$.
	
In addition to the input signals, both the encoder and the decoder can access common randomness (CR), which is a source of randomness involved in the encoding and decoding function, but independent of the semantic source. 
	This may correspond to the stochastic encoder in which the encoder's input includes both indirect observation and a source of randomness. The source of randomness can be the result of a randomized quantization process included in the encoding process for compressing the indirect observation. 
	Although the source of common randomness is not a costly resource in practical compression scenarios,
	we characterize the amount of common randomness as a random integer distributed uniformly over $[2^{nR_c}]$, which is useful for the theoretical analysis. Specifically, the common randomness can be realized by sharing some seeds of the pseduo-random number generator in advance. 
	
	
	
	\subsection{Problem Formulation}
	
The formal definition of an encoder and decoder is as follows:
	\begin{definition}
		For an arbitrary set $\mathcal{W}^n$, a (stochastic) encoder is a function
		\begin{eqnarray}
			{f}:\mathcal{W}^n\times\{1,2,...,2^{nR_c}\}\to \mathcal{Z}^k,
		\end{eqnarray}
		and a (stochastic) decoder is a function
		\begin{eqnarray}
			{g}:\hat{\mathcal{Z}}^k\times\{1,2,...,2^{nR_c}\}\to \hat{\mathcal{W}}^n .
		\end{eqnarray}   
	\end{definition}
    For simplicity, we omit the superscript $m$, and define reconstruction of $W$ under delay $k$ as $\hat{W}_k$
    Then we can minimize the following objective function given $k$:
     \begin{equation}
        \mathcal{L}_k (f,g) =  d_{P}\left(P_{W}, P_{\hat{W}_k}\right)+\lambda\mathbb{E}_{W, g}[d(W, \hat{W}_k)],
        \end{equation}
    where $d_{P}\left(P_{W}, P_{\hat{W}_k}\right)$ is any perceptual measure and $d(W, \hat{W}_k)$ is any distortion measure.
   Note that in our communication goal not only minimizing distortion is needed to achieve pixel-level reconstruction in the classical scheme, but also semantic-level reconstruction by minimizing perception
Naturally, given maximum latency $K$, we should minimize:
    \begin{equation}
        \mathcal{L}_{\text{total}}(f,g) =\sum_{k=1}^{K} \omega_k\mathcal{L}_k,
        \end{equation}
where $\omega_k \geq 0 $.

\section{RSC For Delay-constrained Communication}
        In this section, we present our RSC scheme, which can realize the distion-perception trade-off, while performing rateless transmission for the delay-constrained transmission. In addition, we introduce the dithered quantization and multi-objective optimization, which plays a key role in the stochastic and rateless characteristic of RSC.
\subsection{Design of RSC}
Our design of RSC is inspired by the recent successful application of deep NNs (DNNs), and autoencoders, in particular, to the problem of distribution-preserving lossy compression \cite{dplc2018}, as well as by the promising initial results in the design of end-to-end communication systems using autoencoder architectures \cite{AE2016} \cite{deepintroduction2017}.
Fig.2 displays a structure of RSC in specific applications, e.g., image transmission. With a slight abuse of notation,  the encoder $f_{\phi}$ parameterized by $\phi $ maps $m$-dimensional input image $w $ to a quantized feature vector $z_q$. Following the encoding operation, $z_q$ will lose the tail information by the tail drop operation, which effectively simulates rateless transmission with a delay of $k$, allowing for being decoded at different rates depending on the channel's state by the decoder $g_{\theta}$ parameterized by $\theta$.
\subsection{Dithered Quantization}
Then dithered quantization, a kind of quantization with additive uniform noise, is applied to the output of the perceptron $y$, i.e., $z_q=\left\lfloor y-u\right\rceil$, where $u \sim \mathcal{U}\left([-\frac{\Delta}{2},\frac{\Delta}{2})^{n}\right)$ and ${\Delta}$ is minimum quantization interval, i.e., $\Delta=\frac{2}{L_q-1}$. 
Accordingly, the decoder will add the same noise before decoding with the assumption that $u$ is available to both the encoder and the decoder as a common randomness. Specifically, we generate noise by using a pseudorandom number generator with the same random seed.  Without considering the channel noise and coding rate changes, the entire process of noise addition and denoising can be expressed as\cite{ziv1985universal}:
\begin{equation}
\left\lfloor y+u\right\rceil-u \sim y+u^{\prime},
\end{equation}
where ${u}^{\prime}$ is another uniform noise independent of ${u}$. That is, adding a noise before quantization and then subtracting a noise at the decoding end is equivalent to adding noise directly. This process maps the quantized discrete distribution into a continuous distribution, effectively protecting the original information distribution during the encoding and decoding process.

       \begin{figure}
		\centering
		\includegraphics[width=0.5\textwidth]{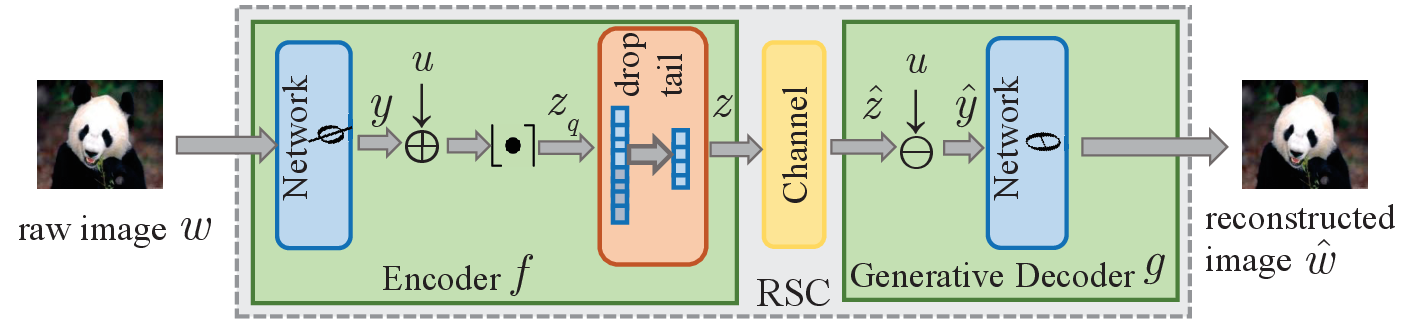}
		\caption{RSC method architecture diagram.}
		\label{f:JSCC_model}
    \vspace{-0.2in}
	\end{figure}
\vspace{-0.1in}
\subsection{ Optimization of RSC}
We next discuss how to optimize (3) further. The distortion between $w$ and the reconstructed image $\hat{w}_k$ can be defined by the MSE as: $d(w,\hat{w}_k)=\|w-\hat{w}_k\|^{2}$. To aid in the assessment and optimization of perceived quality, we adopt the Wasserstein distance, which is the minimum cost of transforming one distribution into another. Specifically, the Wasserstein distance between the source distribution $p_{w}$ and the reconstruction distribution $p_{\hat{w}_k}$ can be approximated by calculating the output of the discriminator in Wasserstein generative adversarial networks (WGAN)\cite{arjovsky2017wasserstein}. Since the optimization goal of the discriminator is to maximize the distance between the two distributions, we evaluate the perceptual quality of our reconstruction by pre-training an optimal discriminator $r_{\zeta}^{\ast}$ parameterized by $\zeta$. Therefore, the perceptual distance in the experiment is defined as follows:
        \begin{equation}
        d_{P}\!\left ( p({\hat{w}_k}),p({w})\right)\!=\!\mathbb{E}_{\hat{w}_k \sim p({\hat{w}_k})}\!\left[r_{\zeta}^{\ast}(\hat{w}_k)\right]\!-\!\mathbb{E}_{w \sim p(w)}\!\left[r_{\zeta}^{\ast}(w)\right].
        \end{equation}
In the proposed RSC, we consider the most widely used additive white Gaussian noise(AWGN) channel in our communication system. Specifically, the symbols encoded by the stochastic encoder will be passed through a fixed additive noise module to simulate channel transmission. This process can be expressed as $\hat{z}=z+n_c$, $n_c\sim\mathcal{N}(0, \sigma^2I_n)$.
Then we can rewrite (3) as:
     \begin{multline}
       {\mathcal{L}}_{k}(f_{\phi},g_{\theta})=\underbrace{\mathbb{E}\left[-\log \left(r_{\zeta}^{\ast}\left(g_{\theta}(f_{\phi}(w;k)+n_c\right), w\right)\right]}_{\text {perception loss }} \\ 
        +  \lambda \underbrace{\mathbb{E}\left[d\left(w, g_{\theta}(f_{\phi}(w;k)+n_c)\right)\right]}_{\text {distortion loss }}.
        \end{multline}
Accordingly, minimizing (4) can be rewritten as:
     \begin{multline}
        {\mathcal{L}_{\text{total}}}(f_{\phi},g_{\theta})\!=\!\sum_{k=1}^{K}\omega_k \bigg\{ {\!\mathbb{E}\!\left[\!-\!\log \left(s_{\zeta}^{\ast}\left(g_{\theta}(f_{\phi}(w;k)\!+\!n_c\right), w\right)\right]} \\ 
        +  \lambda {\mathbb{E}\left[d\left(w, g_{\theta}(f_{\phi}(w;k)+n_c)\right)\right]}\bigg\}.
        \end{multline}

        Thus far, we have developed a new rateless stochastic coding scheme (RSC), which is characterized by its adaptability to channel conditions without a fixed rate, as well as the stochasticity due to shared common randomness.
In the next part, we will conduct a detailed simulation and analysis of RSC.

\section{Experimental Results}
\subsection{Simulation Setup}

To demonstrate the superiority of our proposed rateless stochastic coding scheme, we train and evaluate the scheme on two standard datasets of MNIST and CIFAR-10. And our model mainly consists of an encoder and a generative decoder. For MINIST, we use a fully-connected five-layer perceptron with Batch Normalization and LeakyReLU activation functions at the encoder. In particular, we use the Tanh function after the last layer. 

At the decoder, the denoised $\hat{z}$ will pass through a fully-connected three-layer perceptron with Batch Normalization and LeakyReLU activation functions, which is followed by a deconvolution module. The deconvolution module contains three transposed convolution layers.

To achieve the rateless characteristic of the proposed encoding scheme, we adopt the tail drop scheme proposed by Koike-Akino et al.\cite{koike2020stochastic}. during the training process. Specifically, the output of the quantizer in the encoder passes through a random rate control unit, which makes the tail information of the sequence lost with variable length by using dropout on $z_q$, i.e., $z=dropout(z_q)$.

The distribution of dropped tails used in the experiment is power cumulative distribution, the same as the method in \cite{koike2020stochastic}. Given the maximum coding rate $K$ and any rate $k$, the probability distribution of the actual drop rate $D=K-k$ satisfies: $Pr(D<{\psi}K)={\psi}^{\beta}$, where $\psi$ is the ratio of practical rate to maximum rate $K$ and $\beta$ is set to 0.67 in all experiments to demonstrate the performance of our model without generality. Accordingly, we allocate weights to the loss function of each rate by ${\omega_k=Pr({D})}$ in (4) and (8). Note that unless otherwise noted, the Rate $K$ in our experiment refers to the dimension of the latent dimension after the dropout operation, and the actual coding rate is $ \frac {R \times \log_{2}{L_q}}{s}$ bits per pixel (bpp) given the image size $s$. In addition, the coding rate control unit requires the ACK signal passed by the receiver to ensure that the sent rateless code is successfully decoded and the sender sends the next codeword.

We mainly compare the proposed rateless stochastic coding(RSC) with AE-based traditional JSCC coding (TC) and SAE-based sparse coding (SC), which are widely used by many JSCC programs. Compared with TC, SC encourages the activation of hidden neural unit parts by adding a KL regularization term to the loss function, which is beneficial to the flexibility of the coding rate. For SC, we use 0.1 sparsity degree as a baseline to reflect the flexibility and robustness of coding. To ensure identity, the latent variable dimension of the SC in the experiment is set to 12, the same as the RSC.

\begin{figure}
  \centering



  \subfloat[]{
      \includegraphics[width=0.4\textwidth]{"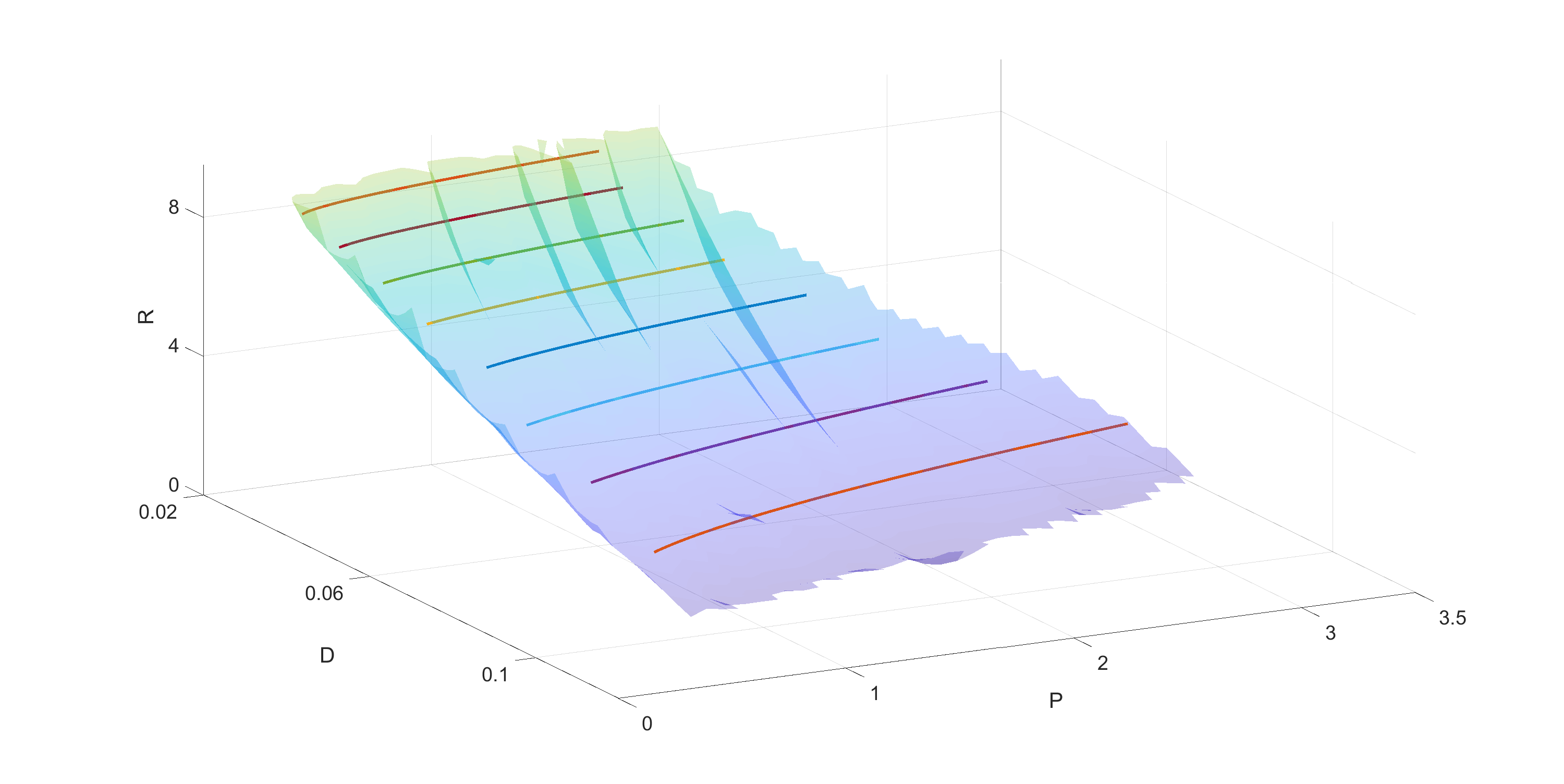"}
      }\hfill
  \subfloat[]{
      \includegraphics[width=0.22\textwidth]{"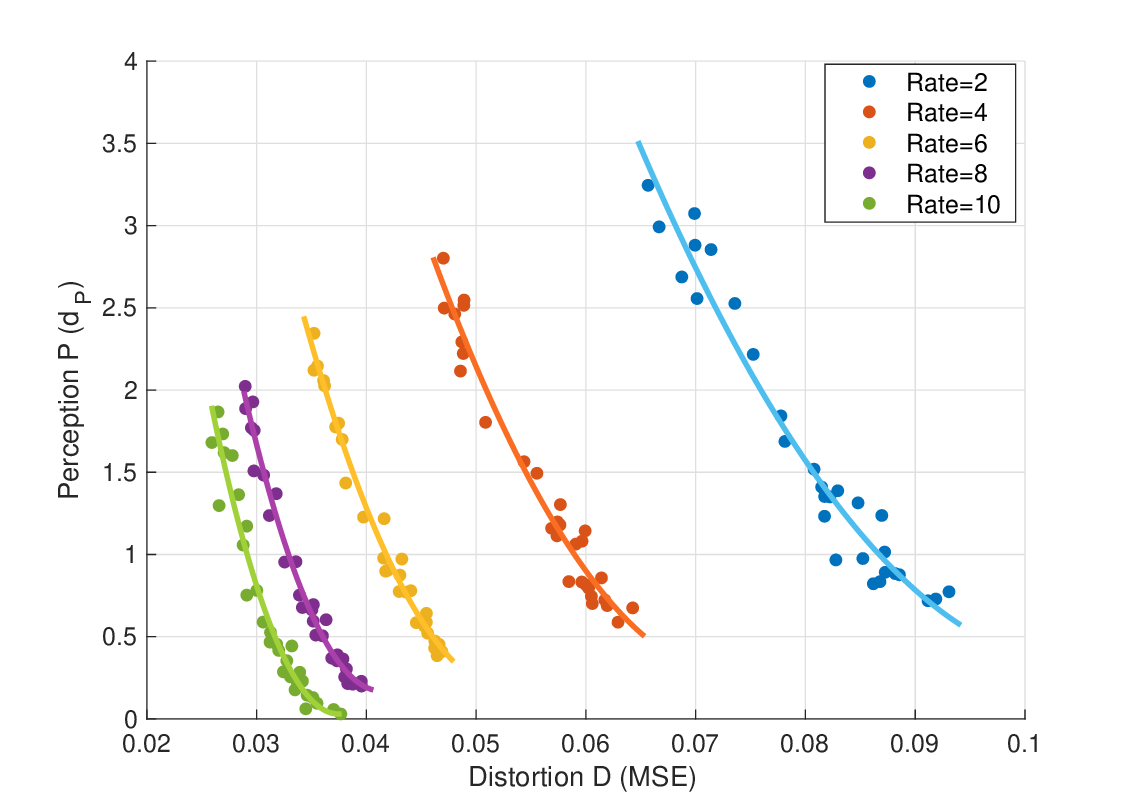"}
      }\hfill
  \subfloat[]{
      \includegraphics[width=0.22\textwidth]{"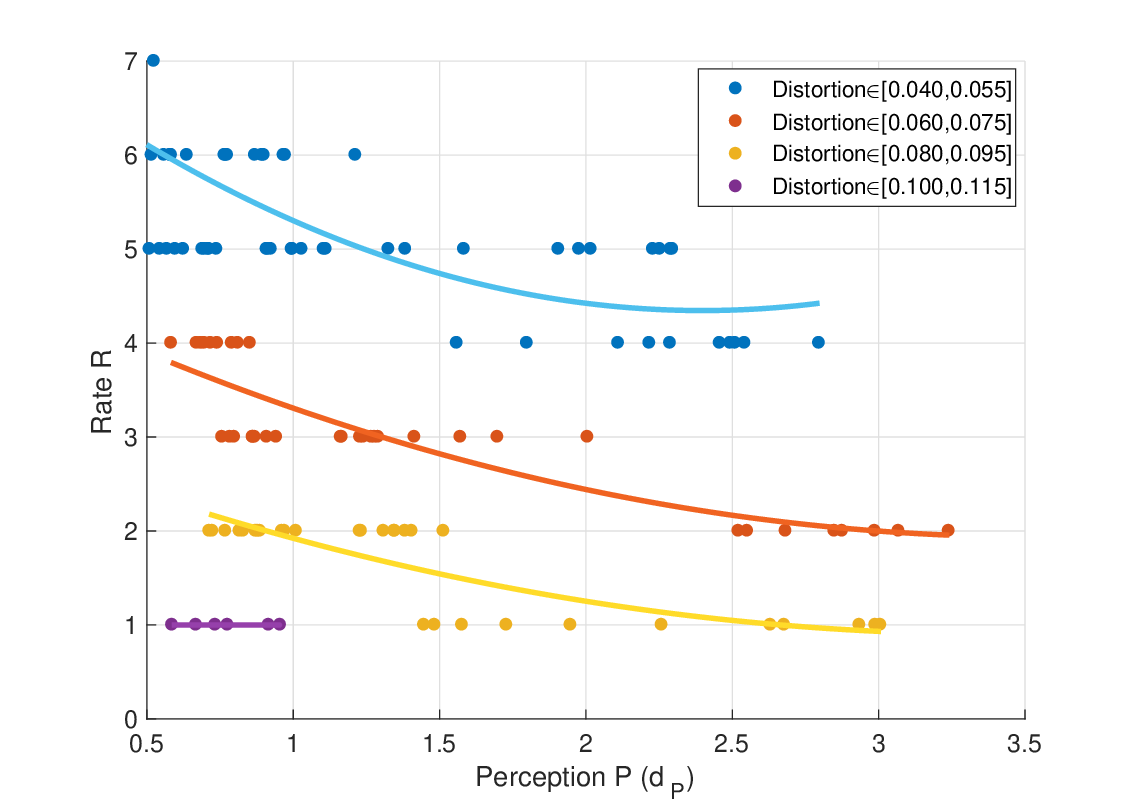"}
      }\hfill
  \caption{The rate-distortion-perception trade-off of MINIST images based on the RSC. (a) Equi-rate lines plotted on the R-D-P trade-off surface highlight the tradeoff between distortion and perceptual quality at different rates. (b) Cross sections of the R-D-P trade-off surface along perception-distortion planes. (c) Cross-sections of the R-D-P trade-off surface along rate-perception planes}
  \label{3figs}
\end{figure}

\subsection{Quantitative Analysis}
 
We first evaluate the rate-distortion-perception(R-D-P) performance of the proposed coding scheme on the MINIST dataset and quantitatively analyze it without considering the channel noise. Then we identify the following key observations:

\begin{observation}
In Fig. 2, we can observe that the R-D-P trade-off as well as its cross sections along the other axis aligned planes of the proposed RSC. In Fig. 2(a),  Equi-rate lines plotted on the R-D-P trade-off surface highlight the trade-off between distortion and perceptual quality at different coding rates.
\end{observation}
\begin{observation}
Fig. 2(b) shows cross sections of the R-D-P trade-off surface along perception-distortion planes. The results show that the proposed RSC can achieve a flexible perception-distortion trade-off at low rates, while changes in perception cause less disturbance to distortion at high bit rates. That is, the proposed RSC can achieve a trade-off between distortion and perception as required at low coding rates. At high coding rates, RSC can achieve good perceptual quality at the expense of small distortion loss. Instead of adjusting the network structure for different rates like common JSCC schemes, the proposed RSC can adaptively change the trade-off strategy at different rates to achieve a more wise reconstruction, which benefits from its rateless characteristics. 
\end{observation}

\begin{observation}
Fig. 2(c) shows the rate-perception trade-off that increasing the rate benefits perceptual quality at constant distortion. Additionally, the slope of the curve indicates that there is significant potential for optimizing perceptual quality under the given coding rate and distortion constraints.
\end{observation}

To confirm that our proposed scheme is an efficient JSCC scheme, we consider examples of image encoding and decoding over additive white Gaussian channels. We first investigate the performance of the proposed RSC by two trade-offs in Fig.3(a), (c), and evaluate its performance under different SNRs in Fig.3 (b), (d). We then make several observations as follows:
\begin{observation}
As can be seen from Fig.3 (a), compared with TC, the curve of distortion with the coding rate of the proposed RSC is convex, while the curve of TC and SC is approximately linear. This shows that in traditional coding schemes, the information contained in each dimension of the compressed and encoded vector is roughly equal, whereas our scheme compresses and encodes vectors with unequal information in different dimensions. Therefore, when reducing coding rates, our scheme tends to preferentially discard data in dimensions that contain less information as a way to achieve better rate-distortion performance, at the cost of only slightly worse performance than the traditional scheme at high coding rates. Note that here we did not introduce perception when evaluating rate-distortion performance to reflect the benefits of adding rateless.
\end{observation}
\begin{figure}
    \centering
    \subfloat[]{
        \includegraphics[width=0.46\linewidth]{"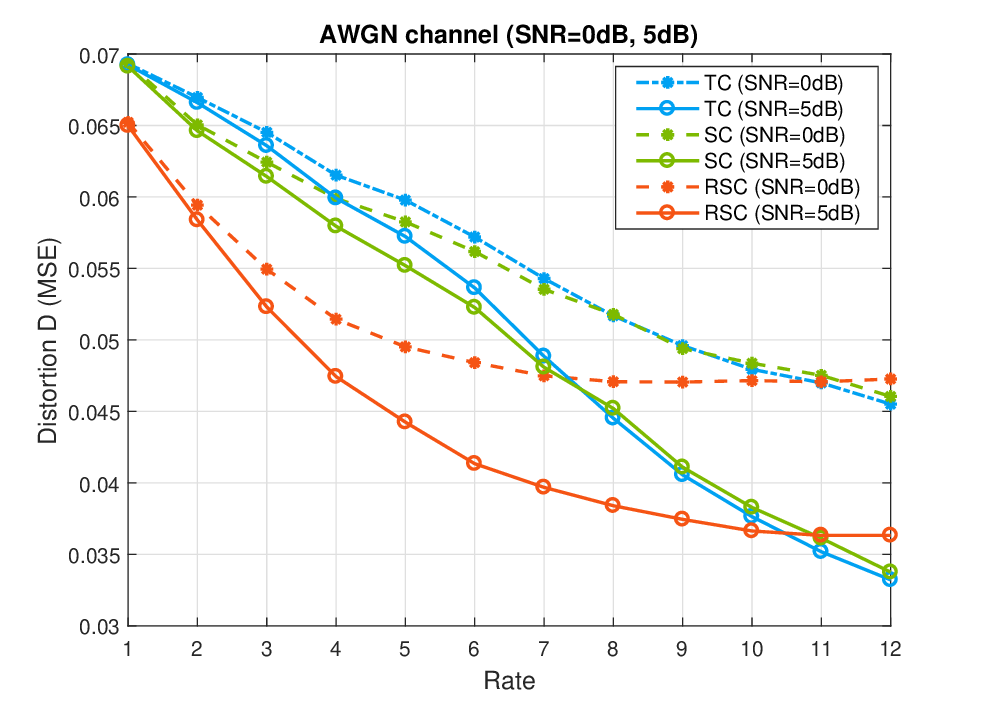"}
        }\hfill
    \subfloat[]{
        \includegraphics[width=0.46\linewidth]{"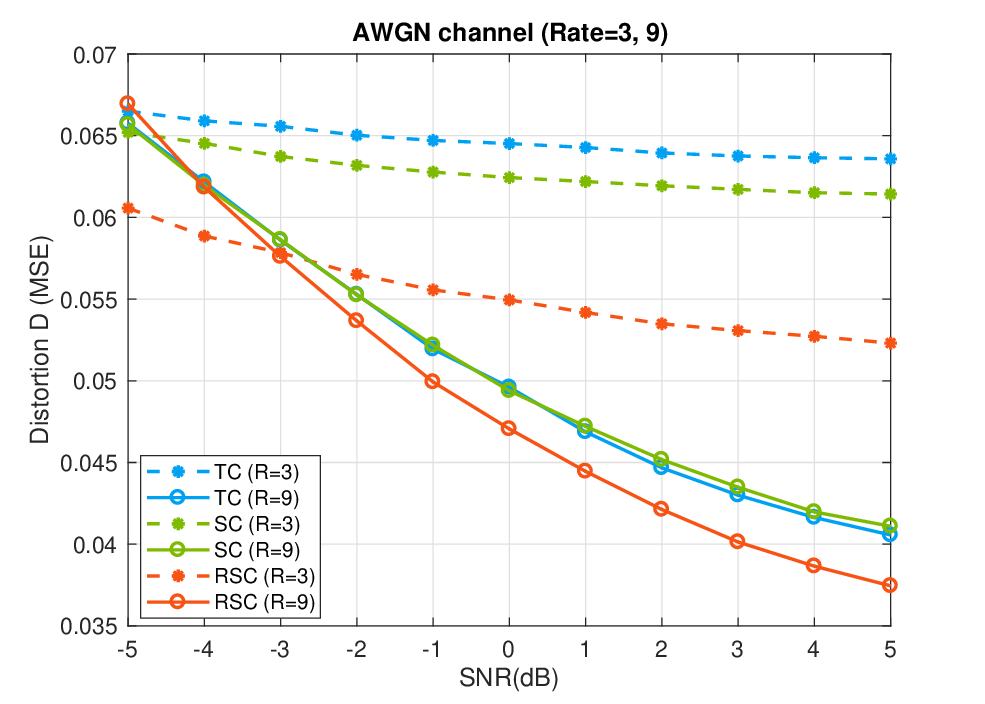"}
        }\hfill
    \subfloat[]{
        \includegraphics[width=0.46\linewidth]{"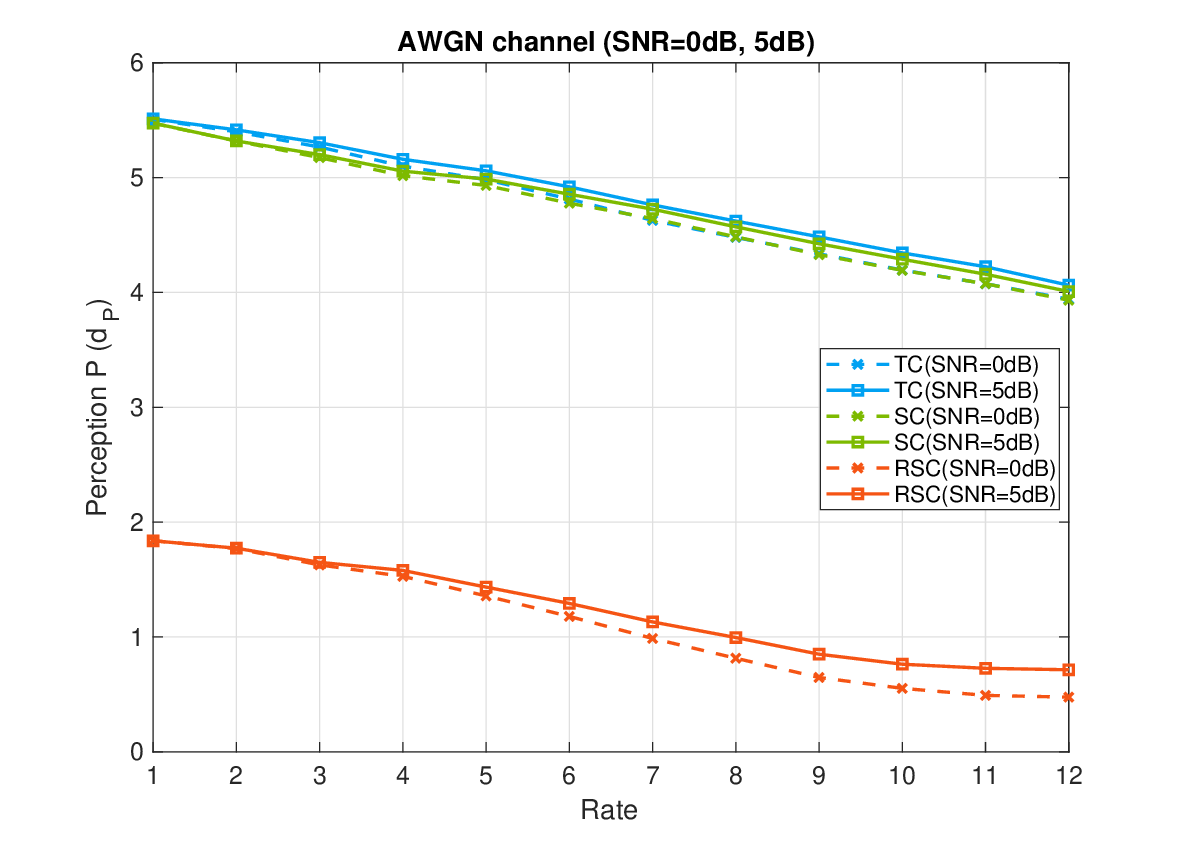"}
        }\hfill
    \subfloat[]{
        \includegraphics[width=0.46\linewidth]{"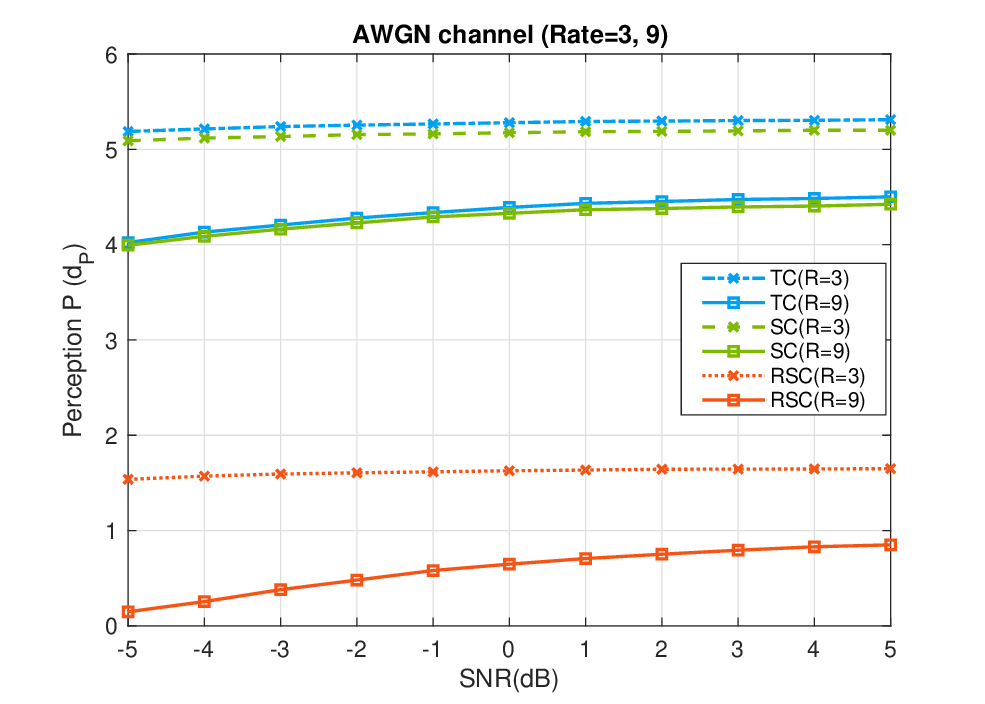"}
        }\hfill
    \caption{Performance comparison of RSC with baseline JSCC schemes on the MINIST dataset over AWGN channels. }
    \label{4figs}
\end{figure}
\begin{observation}
Fig. 3(a), (b) show that our scheme can achieve lower distortion at most rates compared against baselines, especially significant at low rates and high SNRs. The result demonstrates that our RSC scheme has better robustness under different channel environments in practical transmission and achieves image reconstruction in terms of lower distortion standards. 
\end{observation}

More importantly, it is difficult for the common  JSCC schemes to optimize distortion at low rates under the constraints of the rate-distortion criterion. That is, low-rate transmission renders traditional distortion metrics, i.e., MSE, ineffective. To verify that the advantages of the proposed RSC scheme are not limited to the distortion metrics, we present the RSC vs baselines perceptual performance and observe as follows:

\begin{observation}
In Fig. 3(c). we evaluate perceptual performance under the fixed SNR. The results show that RSC can stably reduce the perceptual distance by more than 60\% at various rates. 
\end{observation}
\begin{observation}
Observing Fig. 3(d), we can analyze the impact of SNR on different schemes. Under different channel SNRs, the proposed RSC also outperforms baselines significantly. We also discovered a counter-intuitive phenomenon that the perception of low-rate reconstruction is not sensitive to the variety of SNRs. This can be explained that low-rate encoding will result in a large amount of information being lost, or even compressed into noise. In this case, SNR has a very weak impact on the compressed information. The generative decoder will adopt a random reconstruction strategy when receiving noise so that the perceptual quality of reconstruction depends more on the performance of the generative decoder rather than the coded information received over the channel.
\end{observation}

\begin{figure}
    \centering

     \subfloat[SC / noiseless]{
        \includegraphics[width=0.3\linewidth]{"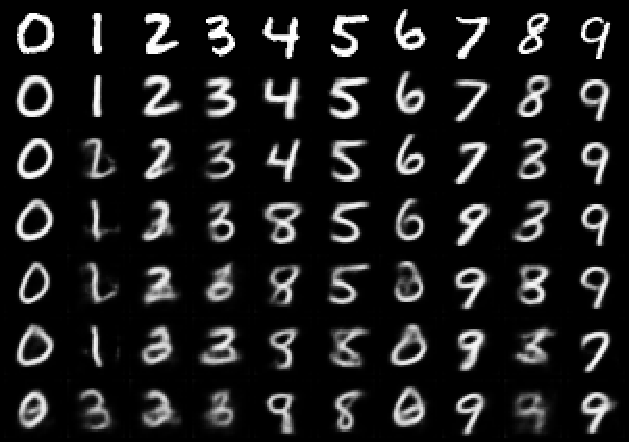"}
        }\hfill
    \subfloat[SC / SNR=5]{
        \includegraphics[width=0.3\linewidth]{"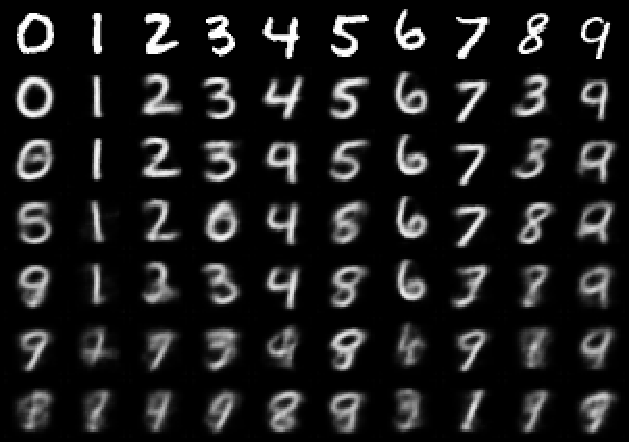"}
        }\hfill
    \subfloat[SC / SNR=0]{
        \includegraphics[width=0.3\linewidth]{"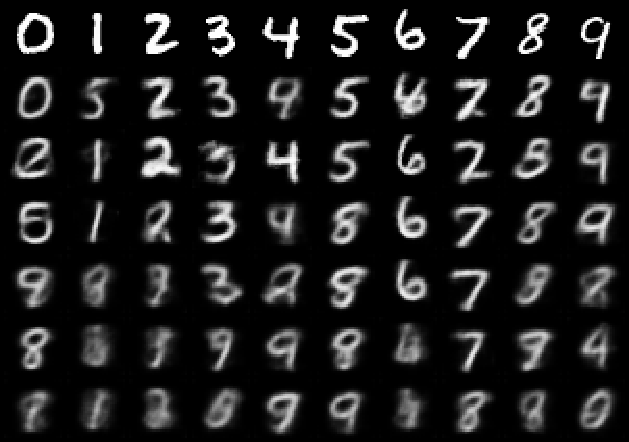"}
        }\hfill
     \subfloat[RSC / noiseless]{
        \includegraphics[width=0.3\linewidth]{"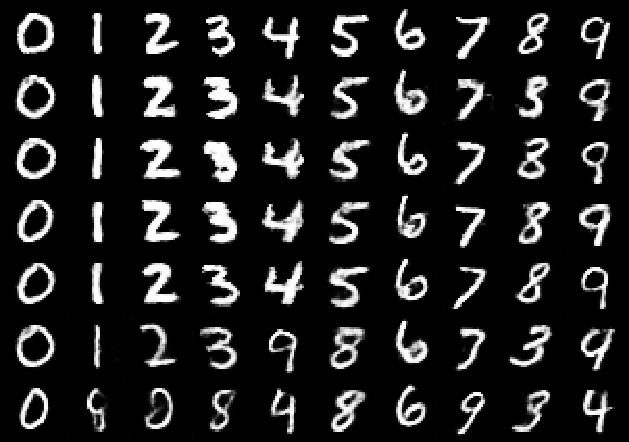"}
        }\hfill
    \subfloat[RSC / SNR=5]{
        \includegraphics[width=0.3\linewidth]{"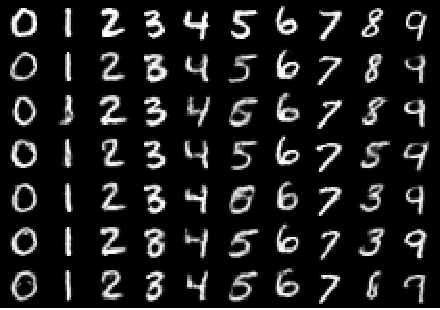"}
        }\hfill
    \subfloat[RSC / SNR=0]{
        \includegraphics[width=0.3\linewidth]{"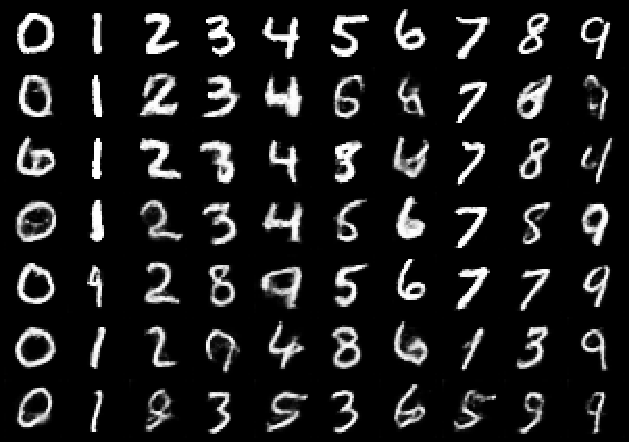"}
        }\hfill
    \subfloat[SC / noiseless]{
        \includegraphics[width=0.3\linewidth]{"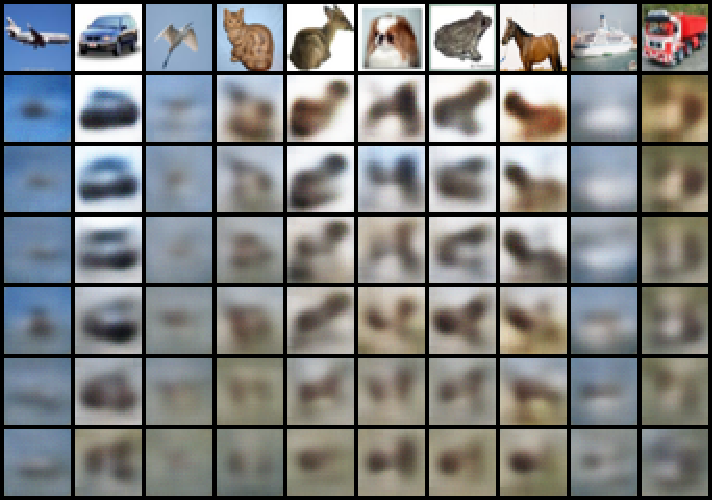"}
        }\hfill
    \subfloat[SC / SNR=5]{
        \includegraphics[width=0.3\linewidth]{"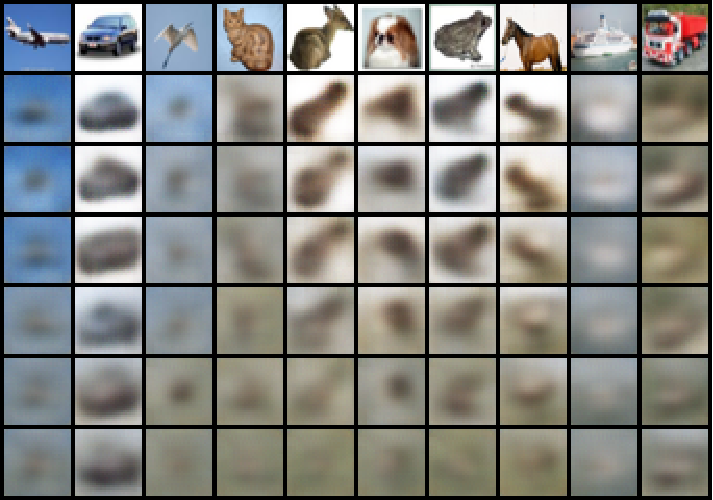"}
        }\hfill
    \subfloat[SC / SNR=0]{
        \includegraphics[width=0.3\linewidth]{"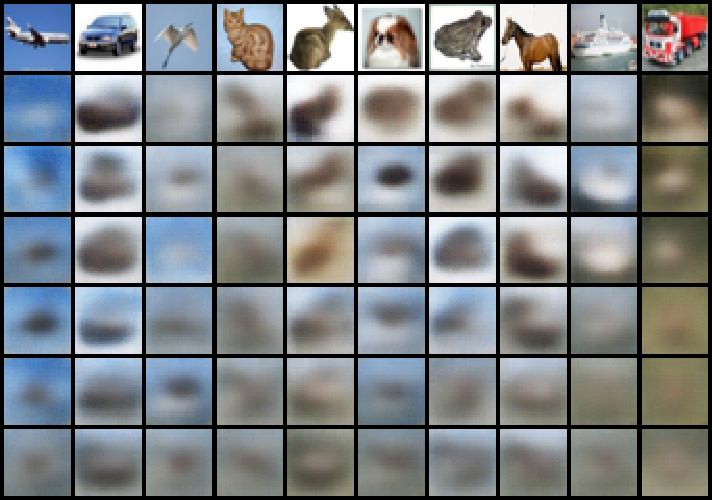"}
        }\hfill 
        
    \subfloat[RSC / noiseless]{
        \includegraphics[width=0.3\linewidth]{"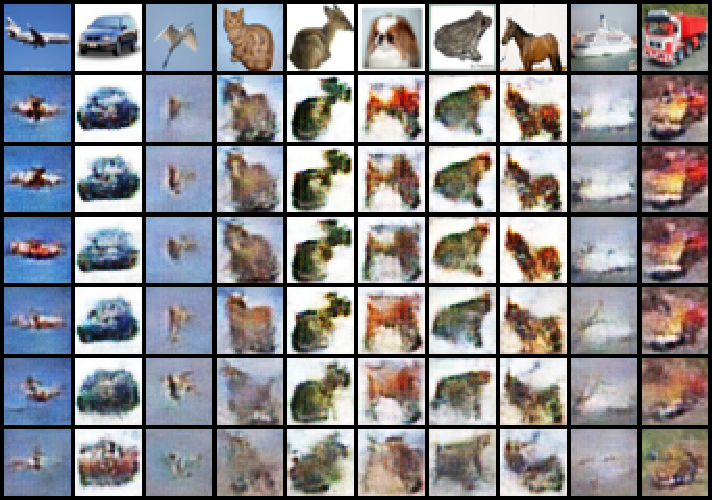"}
        }\hfill
    \subfloat[RSC / SNR=5]{
        \includegraphics[width=0.3\linewidth]{"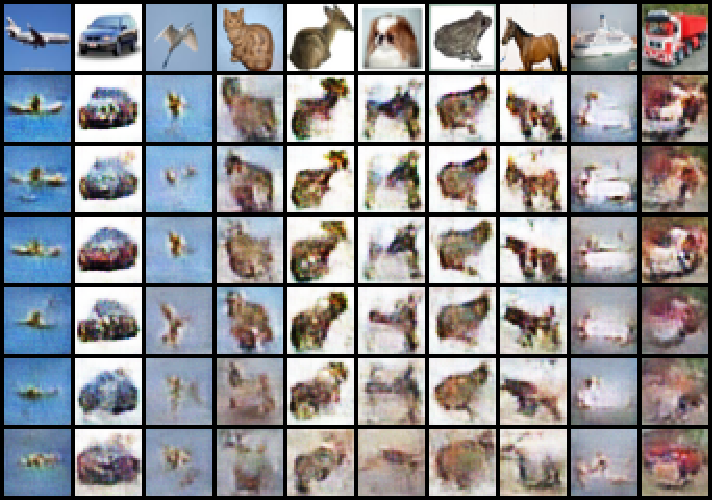"}
        }\hfill
    \subfloat[RSC / SNR=0]{
        \includegraphics[width=0.3\linewidth]{"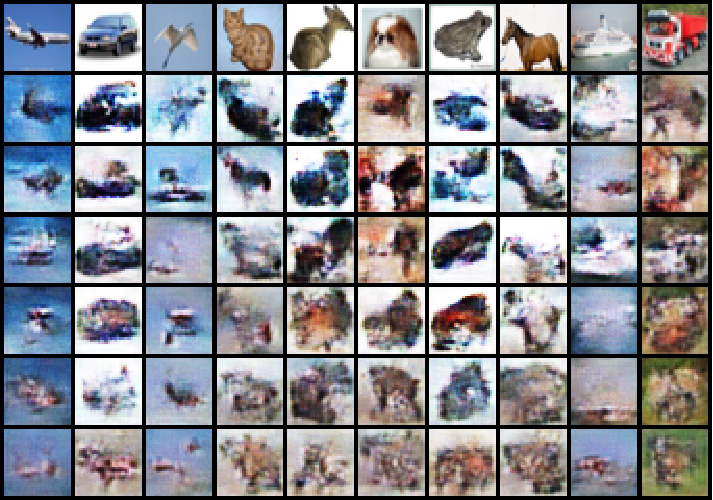"}
        }\hfill
    \caption{Examples of reconstructed images from baselines and the proposed RSC on MNIST and CIFAR-10 varying the survivor rate. The top row of each reconstruction snapshot is the original image, and subsequent rows are reconstructed images for a reduced rate = \{12, 10, 8, 6, 4, 2\}.}
    \label{3figs}
\end{figure}

\subsection{Qualitative Comparison}
We present qualitative comparisons with SC in Fig. 4 and have the following observation:

\begin{observation}
As the rate decreases, the image reconstructed by our scheme is less affected. Especially at lower bit rates, our scheme still achieves high-fidelity image reconstruction well, while SC suffers from more severe blur. The remarkable performance validates that the proposed encoding and decoding scheme effectively realizes the perceptual reconstruction of image semantics at different rates from a subjective perspective intuition. However, we have to admit that there are some errors in the reconstruction, i.e., when SNR = 5 and rate = 8, number 8 is reconstructed as 5 wrongly. It is undeniable that the traditional JSCC scheme with the goal of distortion optimization fails at low rates. Our scheme reconstructs more natural and realistic pictures in this case while allowing some errors. This interesting characteristic can be explained by the fact that our proposed solution can optimize distortion while optimizing perceptual distance, and achieve flexible coding rate strategies during practical transmission. Furthermore, as the SNR decreases, the uncertainty of the proposed RSC reconstruction increases, This indicates that excessive noise makes our scheme tend to prefer images that conform to visual perception rather than reconstructing meaningless blurred images. Surprisingly, even in the case of a harsh channel environment (SNR = 0), a more perfect reconstruction is still achieved at the rate $=$ 5.
\end{observation}
\section{Conclusions}
In this paper, we consider a semantic communication system with CR. 
In contrast to conventional methodologies that solely prioritizes the minimization of distortion, we propose a new communication objective which entails the incorporation of a perceptual component into the objective, thereby facilitating the reconstruction of semantic information at the semantic level.
 Then we propose rateless stochastic coding (RSC) based on the aforementioned theory, a novel delay-constrained coding scheme that enables rateless while achieving a favorable distortion-perception trade-off. At last, the effectiveness of the proposed RSC was verified by simulating practical communication systems, showing it combines the benefits of stochastic coding and rateless coding. 
 The RSC allowed receivers to obtain essentially artifact-free reconstructions at all rates, maintaining the distribution of original data from learning a generative decoder at low rates or low SNRs on the one hand, learning a decoding strategy with almost perfect reconstruction while considering distortion and perception at high rates or high SNRs on the other hand. That is, the proposed RSC is robust to rate reduction and noise in the AWGN channel.

\bibliographystyle{IEEEtran}
\bibliography{main}

\begin{thebibliography}{10}
\providecommand{\url}[1]{#1}
\csname url@samestyle\endcsname
\providecommand{\newblock}{\relax}
\providecommand{\bibinfo}[2]{#2}
\providecommand{\BIBentrySTDinterwordspacing}{\spaceskip=0pt\relax}
\providecommand{\BIBentryALTinterwordstretchfactor}{4}
\providecommand{\BIBentryALTinterwordspacing}{\spaceskip=\fontdimen2\font plus
\BIBentryALTinterwordstretchfactor\fontdimen3\font minus \fontdimen4\font\relax}
\providecommand{\BIBforeignlanguage}[2]{{%
\expandafter\ifx\csname l@#1\endcsname\relax
\typeout{** WARNING: IEEEtran.bst: No hyphenation pattern has been}%
\typeout{** loaded for the language `#1'. Using the pattern for}%
\typeout{** the default language instead.}%
\else
\language=\csname l@#1\endcsname
\fi
#2}}
\providecommand{\BIBdecl}{\relax}
\BIBdecl

\bibitem{shi2021semantic}
G.~Shi, Y.~Xiao, Y.~Li, and X.~Xie, ``From semantic communication to semantic-aware networking: Model, architecture, and open problems,'' \emph{IEEE Communications Magazine}, vol.~59, no.~8, pp. 44--50, 2021.

\bibitem{weaver1953recent}
W.~Weaver, ``Recent contributions to the mathematical theory of communication,'' \emph{ETC: a review of general semantics}, pp. 261--281, 1949.

\bibitem{sesia2004incremental}
S.~Sesia, G.~Caire, and G.~Vivier, ``Incremental redundancy hybrid arq schemes based on low-density parity-check codes,'' \emph{IEEE Transactions on Communications}, vol.~52, no.~8, pp. 1311--1321, 2004.

\bibitem{castura2007rateless}
J.~Castura and Y.~Mao, ``Rateless coding and relay networks,'' \emph{IEEE Signal Processing Magazine}, vol.~24, no.~5, pp. 27--35, 2007.

\bibitem{blau2019rethinking}
Y.~Blau and T.~Michaeli, ``Rethinking lossy compression: The rate-distortion-perception tradeoff,'' in \emph{International Conference on Machine Learning}.\hskip 1em plus 0.5em minus 0.4em\relax PMLR, 2019, pp. 675--685.

\bibitem{zhang2021universal}
G.~Zhang, J.~Qian, J.~Chen, and A.~Khisti, ``Universal rate-distortion-perception representations for lossy compression,'' \emph{Advances in Neural Information Processing Systems}, vol.~34, pp. 11\,517--11\,529, 2021.

\bibitem{blau2018perception}
Y.~Blau and T.~Michaeli, ``The perception-distortion tradeoff,'' in \emph{Proceedings of the IEEE conference on computer vision and pattern recognition}, 2018, pp. 6228--6237.

\bibitem{theis2021coding}
L.~Theis and A.~B. Wagner, ``A coding theorem for the rate-distortion-perception function,'' in \emph{Neural Compression: From Information Theory to Applications--Workshop@ ICLR 2021}, 2021.

\bibitem{theis2021advantages}
L.~Theis and E.~Agustsson, ``On the advantages of stochastic encoders,'' in \emph{Neural Compression: From Information Theory to Applications--Workshop@ ICLR 2021}, 2021.

\bibitem{wagner2022rate}
A.~B. Wagner, ``The rate-distortion-perception tradeoff: The role of common randomness,'' \emph{arXiv preprint arXiv:2202.04147}, 2022.

\bibitem{luby2002lt}
M.~Luby, ``Lt codes,'' in \emph{The 43rd Annual IEEE Symposium on Foundations of Computer Science, 2002. Proceedings.}\hskip 1em plus 0.5em minus 0.4em\relax IEEE Computer Society, 2002, pp. 271--271.

\bibitem{shokrollahi2006raptor}
A.~Shokrollahi, ``Raptor codes,'' \emph{IEEE transactions on information theory}, vol.~52, no.~6, pp. 2551--2567, 2006.

\bibitem{palanki2004rateless}
R.~Palanki and J.~S. Yedidia, ``Rateless codes on noisy channels,'' in \emph{International Symposium onInformation Theory, 2004. ISIT 2004. Proceedings.}\hskip 1em plus 0.5em minus 0.4em\relax IEEE, 2004, p.~37.

\bibitem{castura2006rateless}
J.~Castura and Y.~Mao, ``Rateless coding over fading channels,'' \emph{IEEE communications letters}, vol.~10, no.~1, pp. 46--48, 2006.

\bibitem{chai2023rate}
J.~Chai, Y.~Xiao, G.~Shi, and S.~Walid, ``Rate-distortion-perception theory for semantic communication,'' in \emph{to appear at 2023 IEEE ICNP Workshop}.\hskip 1em plus 0.5em minus 0.4em\relax IEEE, 2023.

\bibitem{dplc2018}
M.~Tschannen, E.~Agustsson, and M.~Lucic, ``Deep generative models for distribution-preserving lossy compression,'' \emph{Advances in neural information processing systems}, vol.~31, 2018.

\bibitem{AE2016}
T.~J. O'Shea, K.~Karra, and T.~C. Clancy, ``Learning to communicate: Channel auto-encoders, domain specific regularizers, and attention,'' in \emph{2016 IEEE International Symposium on Signal Processing and Information Technology (ISSPIT)}.\hskip 1em plus 0.5em minus 0.4em\relax IEEE, 2016, pp. 223--228.

\bibitem{deepintroduction2017}
T.~O’shea and J.~Hoydis, ``An introduction to deep learning for the physical layer,'' \emph{IEEE Transactions on Cognitive Communications and Networking}, vol.~3, no.~4, pp. 563--575, 2017.

\bibitem{ziv1985universal}
J.~Ziv, ``On universal quantization,'' \emph{IEEE Transactions on Information Theory}, vol.~31, no.~3, pp. 344--347, 1985.

\bibitem{arjovsky2017wasserstein}
M.~Arjovsky, S.~Chintala, and L.~Bottou, ``Wasserstein generative adversarial networks,'' in \emph{International conference on machine learning}.\hskip 1em plus 0.5em minus 0.4em\relax PMLR, 2017, pp. 214--223.

\bibitem{koike2020stochastic}
T.~Koike-Akino and Y.~Wang, ``Stochastic bottleneck: Rateless auto-encoder for flexible dimensionality reduction,'' in \emph{2020 IEEE International Symposium on Information Theory (ISIT)}.\hskip 1em plus 0.5em minus 0.4em\relax IEEE, 2020, pp. 2735--2740.

\end{thebibliography}


\end{document}